## Analytical model for Stirling cycle machine design

### *Corresponding author*

Fabien Formosa, Laboratoire SYMME - Université de Savoie

BP 80439- 74944 ANNECY LE VIEUX CEDEX - FRANCE

Tel: +33 4 50 09 65 08

Fax: +33 4 50 09 65 43

Mail: fabien.formosa@univ-savoie.fr

### *Authors*

F. Formosa[1], G. Despesse[2]

### *Affiliations*

[1] Laboratoire SYMME, Université de Savoie, Annecy le Vieux, France

[2] Laboratoire Capteurs Actionneurs et Récupération d'Energie, CEA-LETI-MINATEC, Grenoble, France

### *Abstract*

In order to study further the promising free piston Stirling engine architecture, there is a need of an analytical thermodynamic model which could be used in a dynamical analysis for preliminary design. To aim at more realistic values, the models have to take into account the heat losses and irreversibilities on the engine. An analytical model which encompasses the critical flaws of the regenerator and furthermore the heat exchangers effectivenesses has been developed. This model has been validated using the whole range of the experimental data available from the General Motor GPU-3 Stirling engine prototype. The effects of the technological and operating parameters on Stirling engine performance have been investigated. In addition to the regenerator influence, the effect of the cooler effectiveness is underlined.

*Keywords*

Stirling engine; Design; Analytical model; Thermal losses; Mechanical losses

## 1. Introduction

The Stirling engine, invented in 1816, running according to a reversible closed cycle knew a practical use as a reliable and sure engine, during almost one century before being supplanted by the spark-ignition engine. Nowadays, the Stirling machines are in commercial use only as heat pump, used mainly for cryogenic cooling and air liquefaction. As an engine, the Stirling remains a field of numerous researches and development works. Recent experimental realizations demonstrate power densities as well as noteworthy efficiencies [1]. One of the most promising applications of the Stirling cycle is the free piston Stirling engine (FPSE) configuration [2-4].

The optimal design of FPSEs is a difficult task. Indeed, no mechanical linkage fixes the strokes and phase angle for the moving elements. Hence, a global dynamic analysis is required to predict the periodic steady operation. Due to the complexity of this analysis, the isothermal assumption is usually adopted. Hence, the pressure as a function of the piston and displacer positions can be expressed in an analytical way. Linearization methods are then used to obtain the performances of the engine [5-8]. However, these models do not take into account the thermal losses of the engine which lead to erroneous predictions of the performances. Therefore, there is a need for an accurate analytical thermodynamic isothermal model which can be used in accordance with the dynamical analysis of the FPSE for preliminary design purpose.

Many investigators have studied the effect of some heat losses and irreversibilities on the engine performance indices. In the many parameters to be taken into account, dead volume as well as non-ideal regeneration have the highest influence on the Stirling performances compare to all the technological parameters of a practical engine [9, 10]. Popescu et al. [11] showed that the most significant reduction in performance is due to the non-adiabatic regenerator. Kongtragool [12] studied the influence of the regenerator efficiency and the dead volumes on the work as well as the efficiency of the machine. However, this study does not include the heat transfers through the temperature difference at the heat source and sink.

On the basis of the conventional entropy techniques, for the studying of solar Stirling engine cycle performance, Costea et al. [13] included the effects of heat transfers, incomplete heat regeneration and irreversibilities of the cycle as conduction, pressure losses or mechanical friction between the moving parts. Timoumi et al. [14] developed a precise second order model which includes all the losses at the same time. The method based on a lumped analysis approach leads to a numerical model and has been used for the optimization of the General Motors GPU-3 [15, 16].

But this type of models appears to be not suitable for a preliminary design stage of FPSE. Nevertheless, the extensive study based on the GPU- is a reference model to validate new developments.

Stirling machine with dead volume can be analytically studied using the Schmidt approach [17] as far as isothermal evolutions and ideal regeneration are assumed. Following a second order approach according to Martini [15] classification, the Schmidt results can be completed by an energetic balance including a non ideal regeneration. In addition to the previous limitations, the thermal transfers between the machine and its surrounding thermal sources have to be taken into account in a Stirling machine model. They induce great thermal losses which have to be limited in technological realizations. Heat exchangers at the hot and cold ends transfer the thermal energy to the working fluid. The necessary thermal gradient to induce this transfer gives way to a temperature difference between the thermal source and the working fluid at the considered part of the machine. The Newton's law allows the modelling of the relation between the thermal transfer rate and the temperature difference. In the work of Senft [18], a constant effectiveness of thermal exchangers with respect to fluid flow rate is assumed. Given that on the one hand, the heat transfers are related to the operating frequency and on the other hand the frequency appears to be a main optimization parameter, the study of Senft must be examined in more details. Consequently, the model developed in this study accounts for the effect of the flow rate to evaluate the thermal effectiveness of the heat exchangers.

The present work consists in the elaboration and the validation of an analytical isothermal model as a first step to the development of an accurate global analysis of FPSE. The obtained analytical model can be used to analyse Stirling engines including their dead volume, imperfect regenerator and external in addition to internal thermal transfers.

Firstly, the analytical Schmidt results are recalled. With the addition of the thermal flaws, energetic characteristics of a Stirling engine are set out. Then, the heat exchangers and the regenerator efficiencies as a function of fluid flow rates are added. Though it is not a free piston type engine, the GPU-3, it is one of the most documented Stirling engine and its data are therefore used to compare the model results to the experimental ones. For a wide range of different pressures and frequencies, the model shows a good correlation with the experimental data. The model is also compared to the classical adiabatic second order results and shows good agreements with them. In the last part, an optimization is performed.

## *2. Thermodynamic analysis*

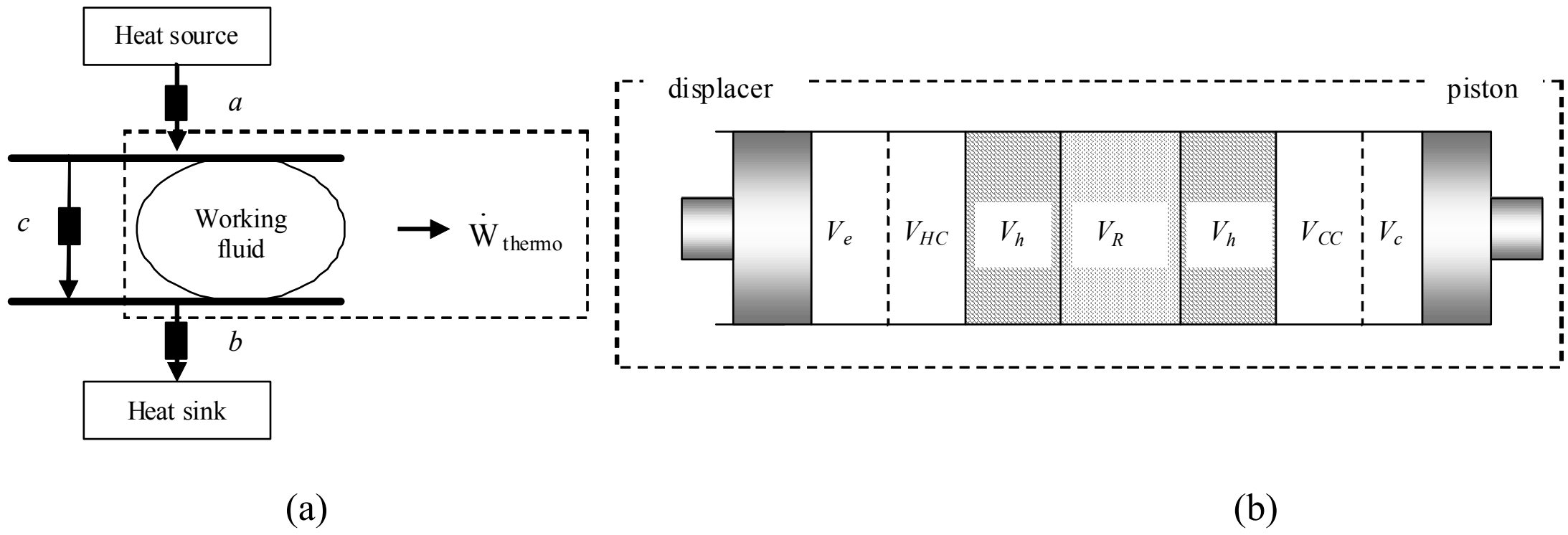

Fig. 1 General schemes of a Stirling machine scheme with thermal interfaces.

In this study, the conventional schematic representation of a “thermodynamic machine” (framed part in Fig. 1 a) is completed by three thermal coefficients. Denoted *a*, *b* and *c*, they are associated with the heat exchangers efficiencies for the first two as well as conduction phenomenon for the latter.

The Newton’s law for expressing heat losses will be adopted to model the thermal transfers in the machine.

## 2.1. Stirling engine characteristics

The generic schemes of a Stirling machine as well as the temperature distribution are described in Fig. 1. The displacer takes place in the left end whereas piston is at the opposite side.

In order to obtain an analytical expression of the engine pressure, the usual following assumptions are adopted:

- Working fluid within the engine chambers stays at constant and unvarying temperatures.
- The temperature of the working fluid within heat exchangers volumes $V_h$ and $V_k$ are $T_U$ and $T_L$ respectively.
- Temperature within the regenerator can be described by a linear evolution between $T_U$ and $T_L$.
- The regenerator behaviour is symmetrical.
- The ideal gas law is adopted.
- The pressure is the same throughout the machine for each considered position of both the piston and the displacer.
- Harmonic movements for the piston and the displacer.

Organ has demonstrated that an equivalent Stirling machine always exists [19]. As a result, an equivalent Stirling machine whatever alpha, beta or gamma Stirling engine mechanical arrangement considered can be assed using the same one-dimensional geometry model. Therefore, we choose here to develop our model on such an equivalent Stirling machine for which expansion and compression volumes evolutions are:

$$\begin{aligned} V_e &= V_d/2\ (1 + \cos \phi(t)) \\ V_c &= \kappa V_d/2\ (1 + \cos (\phi(t) - \alpha)) \end{aligned} \qquad (1)$$

In which $\phi(t)$ is the angle position and $\alpha$ the phase angle. The piston and displacer swept volume ratio is denoted $\kappa = V_p / V_d$.

The total mass of gas inside the various chambers is defined as the sum of the mass of the gas contained in each of the chamber defined in Fig. 2:

$$m = m_{HC} + m_h + m_R + m_i + m_k + m_{CC} \quad (2)$$

If we substitute the expressions of the masses in (2) with the ideal gas law (e.g. $m_{HC} = \frac{p\, V_{HC}}{R\, Th}$), we obtain:

$$p = m\, R \,/\, \left(\frac{V_e}{T_U} + \frac{V_h}{T_U} + \frac{V_R}{T_R} + \frac{V_k}{T_L} + \frac{V_c}{T_L}\right) \quad (3)$$

In which $R$ is the considered gas constant per unit of mass and $T_R$ the mean temperature of the regenerator.

The Schmidt analysis results [17] are recalled hereafter:

$$p = \frac{m\, R}{s} \frac{1}{1 + \beta \cos(\phi(t) - \theta)} \quad (4)$$

With:

$$\beta = \frac{\sqrt{2\, \tau \kappa \cos \alpha + \tau^2 + \kappa^2}}{2\, \nu + \kappa + \tau}$$

$$\tan \theta = \frac{\kappa \sin \alpha}{\kappa \cos \alpha + \tau}$$

$$\frac{1}{s} = 2 \frac{T_U}{V_d} \tau \frac{1}{2\, \nu + \kappa + \tau} \quad (5)$$

$$\nu = \frac{V_{CC}}{V_d} + \frac{V_R}{V_d}\frac{T_L}{T_R} + \frac{V_{HC}}{V_d} \tau$$

In which $\tau = T_L / T_U$.

Consequently, the main characteristic parameters of the engine are given by the following analytic expressions:

- Mean effective pressure

$$p_{mean} = \frac{m\,R}{s} \frac{1}{\sqrt{1-\beta^2}} \tag{6}$$

- Heat added (*Qe*) and heat rejected (*Qc*) for each cycle:

$$Qe = \int_{\text{cycle}} p\,\mathrm{d}V_{HC} = p_{mean} \frac{\pi\,V_d}{\beta} \left(1 - \frac{1}{\sqrt{1-\beta^2}}\right) \sin\theta \tag{7}$$

$$Qc = \int_{\text{cycle}} p\,\mathrm{d}V_{CC} = -\,p_{mean} \frac{\pi\,\kappa\,V_d}{\beta} \left(1 - \frac{1}{\sqrt{1-\beta^2}}\right) \sin(\alpha-\theta) \tag{8}$$

*Qe* stands for the heat added to the “hot part” of the engine and *Qc* for the “cold part”.

- Efficiency and power

The ideal thermodynamic efficiency is exactly the Carnot efficiency:

$$\eta_i = 1 - \frac{T_L}{T_U} \tag{9}$$

Useful mechanical work can be determined from equations (7) and (8) and varies notably as a function of the mean effective pressure:

$$W_i = p_{mean}\,V_d \frac{\pi\,\kappa\,(\tau-1)\sin\alpha}{\beta\sqrt{\kappa^2+\tau^2+2\,\kappa\,\tau\cos\alpha}} \left(1-\sqrt{1-\beta^2}\right) \tag{10}$$

Thus, the indicated power can be expressed as the product of the mechanical work by the operating frequency *f*:

$$P_i = f\,W_i \tag{11}$$

## 2.2. Thermal approach

- Thermal power

The cycle average power turns out to be given in an alternative way. According to the power balance of the machine as described as in Fig. 1 a, the cycle average power can expressed as:

$$P_{th} = a\, T_H\, (1 + \delta\, \Gamma - \xi - \xi\, \delta\, \tau) \qquad (12)$$

In which $\xi = T_U / T_H$ is the temperature ratio between the heat source and the highest temperature of the engine, $\delta = b/a$ and $\Gamma = T_C / T_H$ .

- Thermal efficiency

The losses associated to the regenerator have the highest influence on the performances. We anticipate that the shuttle losses as well as the gas spring hysteresis have little influence on the performances. By doing this, these losses are neglected compared to the former ones.

The thermal efficiency of the engine is defined by the ratio of the available power by the added thermal power:

$$\eta_{th} = \frac{P_{th}}{\dot{Q}_h + \dot{Q}_T + \dot{Q}_R} \qquad (13)$$

In which $\dot{Q}_T$ is the conduction loss, and $\dot{Q}_R$ is the thermal power related to the regenerator inefficiency.

We define the various thermal impedances ratios:

$$\rho_{cond} = c_{cond} / a \qquad (14)$$

$$\rho_R = \dot{m}_R\, C_v / a \qquad (15)$$

Where $\dot{m}_R$ is the fluid mass rate inside the regenerator.

Hence:

$$\dot{Q}_T = a\, \rho_{cond}\, T_U\, (1-\tau) \qquad (16)$$

$$\dot{Q}_R = (1-e)\, a\, \rho_R\, T_U\, (1-\tau) \qquad (17)$$

The regenerator effectiveness $e$ depends on the temperatures of the engine. We define $e = \frac{T'_U - T_L}{T_U - T_L}$ with $T'_U < T_U$ the highest actual outlet temperature of the fluid from the regenerator. It must be noted that as far as a symmetrical regenerator behaviour is assumed a single effectiveness is defined and $e$ can be conversely defined as $e = \frac{T'_L - T_U}{T_L - T_U}$ with $T'_L > T_L$ the actual lowest temperature of the fluid from the regenerator.

Finally the thermal efficiency is expressed as a function of the non-dimensional parameters:

$$\eta_{th} = \frac{1 + \delta\Gamma - \xi - \xi\tau\delta}{1 - \xi + (\rho_{cond} + (1-e)\rho_R)\xi(1-\tau)} \qquad (18)$$

- Thermodynamic conditions

The second law of thermodynamics requires that the thermal efficiency does not exceed the Carnot efficiency that is:

$$\eta_{th} \leq 1 - \tau \qquad (19)$$

For a machine which operates as an engine the available power $P_{th}$ is always positive. Then we set:

$$P_{th} \geq 0 \qquad (20)$$

Therefore, using the two previous conditions (19) and (20), the permissible values of $\xi$ are within an hatched domain shown on Fig. 2.

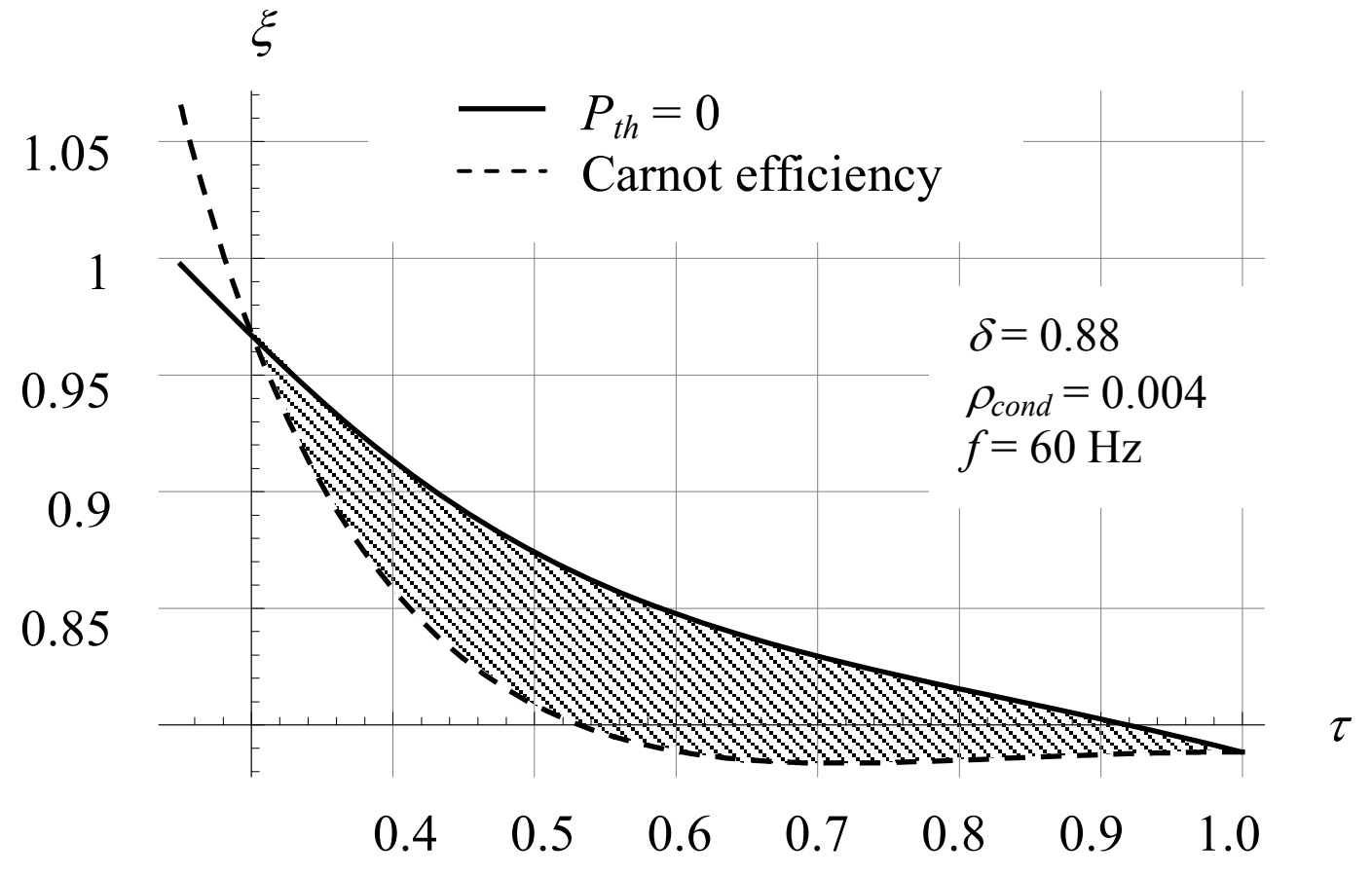

Fig. 2. Permissible area for $\xi$.

In order to obtain the maximum performances an optimal case is chosen. Thus, the inequality (19) can be switched to equality. Consequently, the optimal ratio $\xi_{optim}$ between the heat source temperature and the temperature of the expansion chamber of the engine can be obtained. Using equation (18):

$$\xi_{optim} = \frac{\delta\,\Gamma + \tau}{(\delta + 1)\ \tau - (\rho_{cond} + (1\text{-}e)\ \rho_R)\ (1-\tau)^2} \tag{21}$$

### 2.2.1. Expression of the thermal coefficients

According to the Newton's law the coefficients $a$ and $b$ are assumed to be representative of the heat exchange for the hot and the cold end respectively. Thermal effectiveness of the heat exchangers are dependant on the machine parameters and more specifically the operating frequency $f$. Therefore, it is important to consider this dependency to deepen to the work of Senft [18] where a constant thermal coefficients are assumed.

The $a$ and $b$ thermal coefficients are deduced from a simple convection model relations:

$$a = h_h\, A_{wh}$$
$$b = h_k\, A_{wk} \tag{22}$$

In which $h_h$ and $h_k$ are the convection heat transfers, $A_{wh}$ and $A_{wk}$ are the wetted areas of the heater. Based on experimental correlations, $h$ is expressed as a function of the Colburn J-factor $J_h$ [20].

$$J_h(Re) = \frac{h\, Pr^{2/3}}{C_p\, \dot{m}/A_{ff}} \tag{23}$$

In which $Pr$ is the Prandtl number and $A_{ff}$ the free flow area and the cooler respectively. $C_p$ is the constant pressure specific heat capacity of the working fluid.

The evolution of $J_h$ with respect to $Re$ is of the form $J_h(Re) = C_1\, Re^{-n}$. For tubular heat exchangers which are used for the GPU-3 Stirling engine, the Colburn J-factor for heat transfer is:

$Re \leq 3\,000 \quad J_h = \exp(0.337\text{-}0.812 \log(Re))$

$3000 \leq Re \leq 4\,000 \quad J_h = 0.0021$

$4\,000 \leq Re \leq 7\,000 \quad J_h = \exp(13.31+0.861 \log(Re))$

$7\,000 \leq Re \leq 10\,000 \quad J_h = 0.0034$

$10\,000 \leq Re \quad J_h = \exp(-3.575\text{-}0.229 \log(Re))$

A polynomial interpolation for $J_h$ will be used in the model.

Let us now move to the Reynolds number $Re$. The mass flow rate can be estimated from de Schmidt analysis and the Organ works [19]. Fig. 3 plots the Reynolds number for the heater and the cooler as a function of $\tau$ for the GPU-3 parameters. The usually reported turbulent behaviour of flow is predicted.

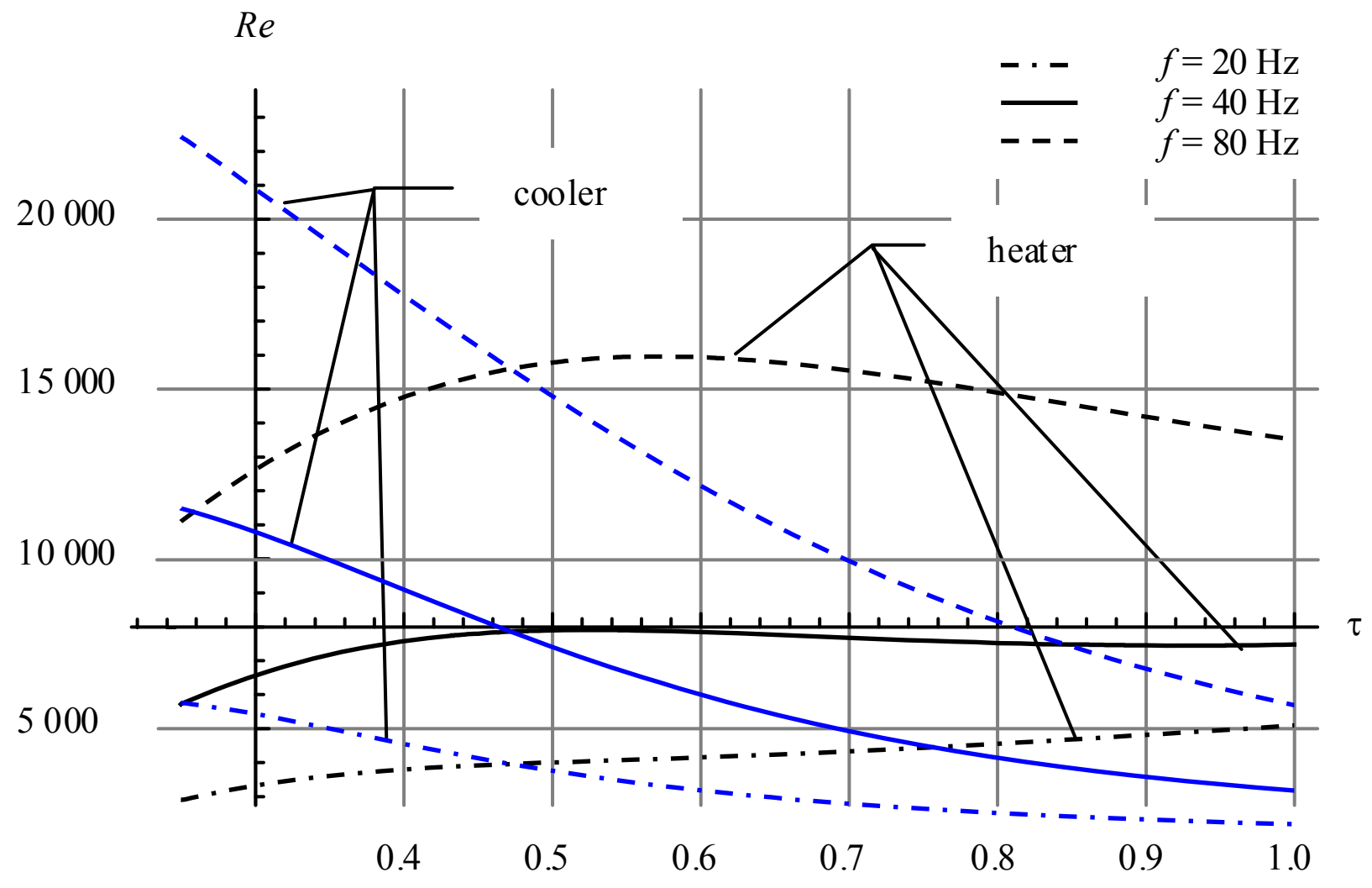

Fig. 3. Evolution of the Reynolds numbers for the heat exchangers with respect to temperature ratio $\tau$ for three operating frequencies.

The thermal coefficients can eventually be evaluated from the engine geometric and operating parameters.

### 2.2.2. Expression of losses

- Reheat loss

A non-ideal regenerator can not provide an outlet fluid temperature equals to the adjacent chamber temperature. It follows that the hot heat exchanger must provide an extra heat called $\dot{Q}_R$ defined by equation (17).

Simplified expressions of $e$ with respect to NTU can be found in the literature. They all involve NTU and some include additional operating parameter such as the flush ratio FR [21].

The equations (24) hereafter give different expressions of $e$:

$e_M = \frac{\text{NTU}}{2+\text{NTU}}$ as proposed by Martini [15]

$e_O = 1 - \frac{1}{\text{NTU}}$ is suggested by Organ [19]

(24)

$e_{DM} = \frac{\text{NTU}}{2+\text{NTU}} + \frac{2\ \text{FR}}{\text{NTU}\ (2+\text{NTU})}$ f(FR) Where f(FR) = 1- exp(-NTU) if FR $\leq$ 1 and

f(FR) = 1 - exp(-NTU/FR) otherwise, by De Monte [21]

The NTU can be evaluated through the correlation of the Colburn factor for a mesh grid regenerator with the Reynolds number [15]. Once again, the calculation of the Reynolds number for the regenerator relies on a coupled Schmidt – Organ approach. As shown in Fig. 4, laminar flows occur in the regenerator for the considered operating frequencies.

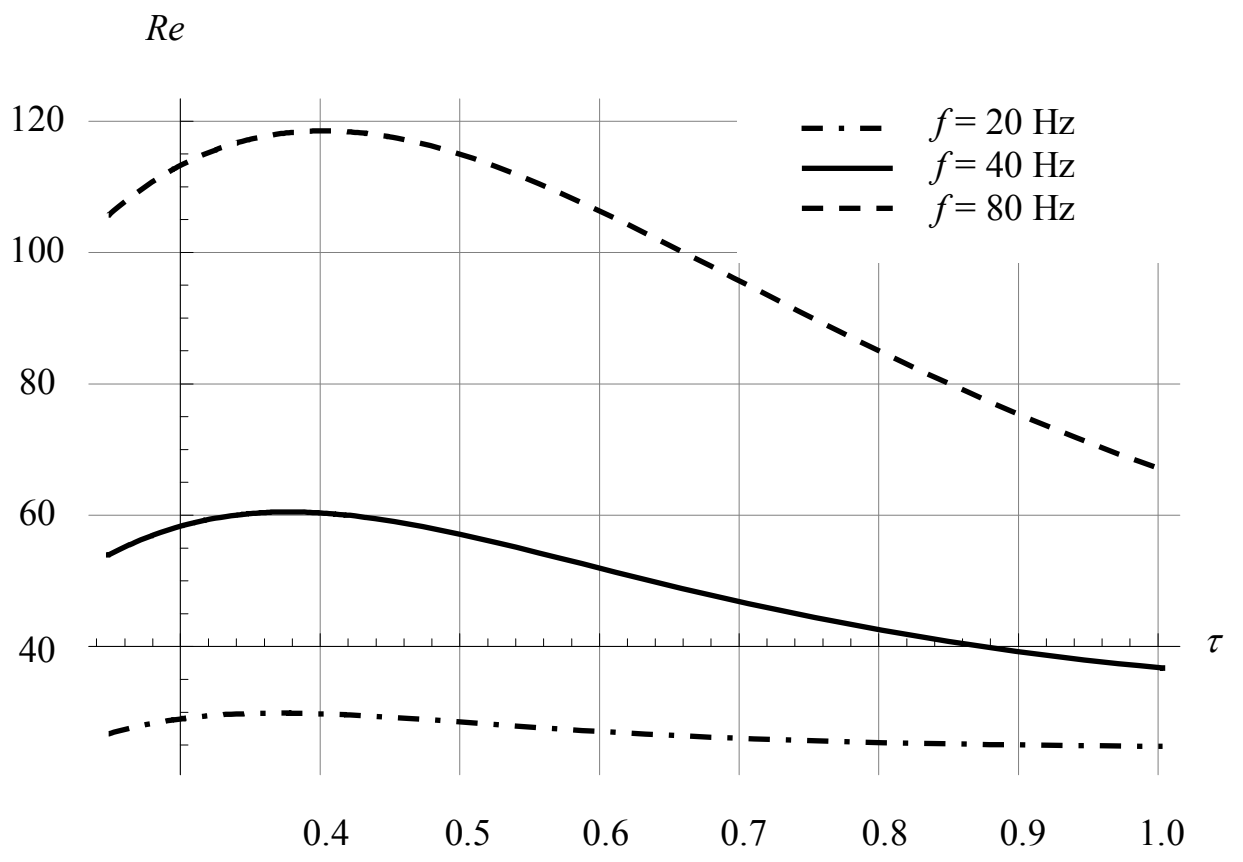

Fig. 4. Evolution of the Reynolds number for the regenerator with respect to temperature ratio $\tau$ for three operating frequencies.

The results for the various expressions of $e$ with respect to the temperature ratio $\tau$ and for three operating frequencies are plotted in Fig. 5.

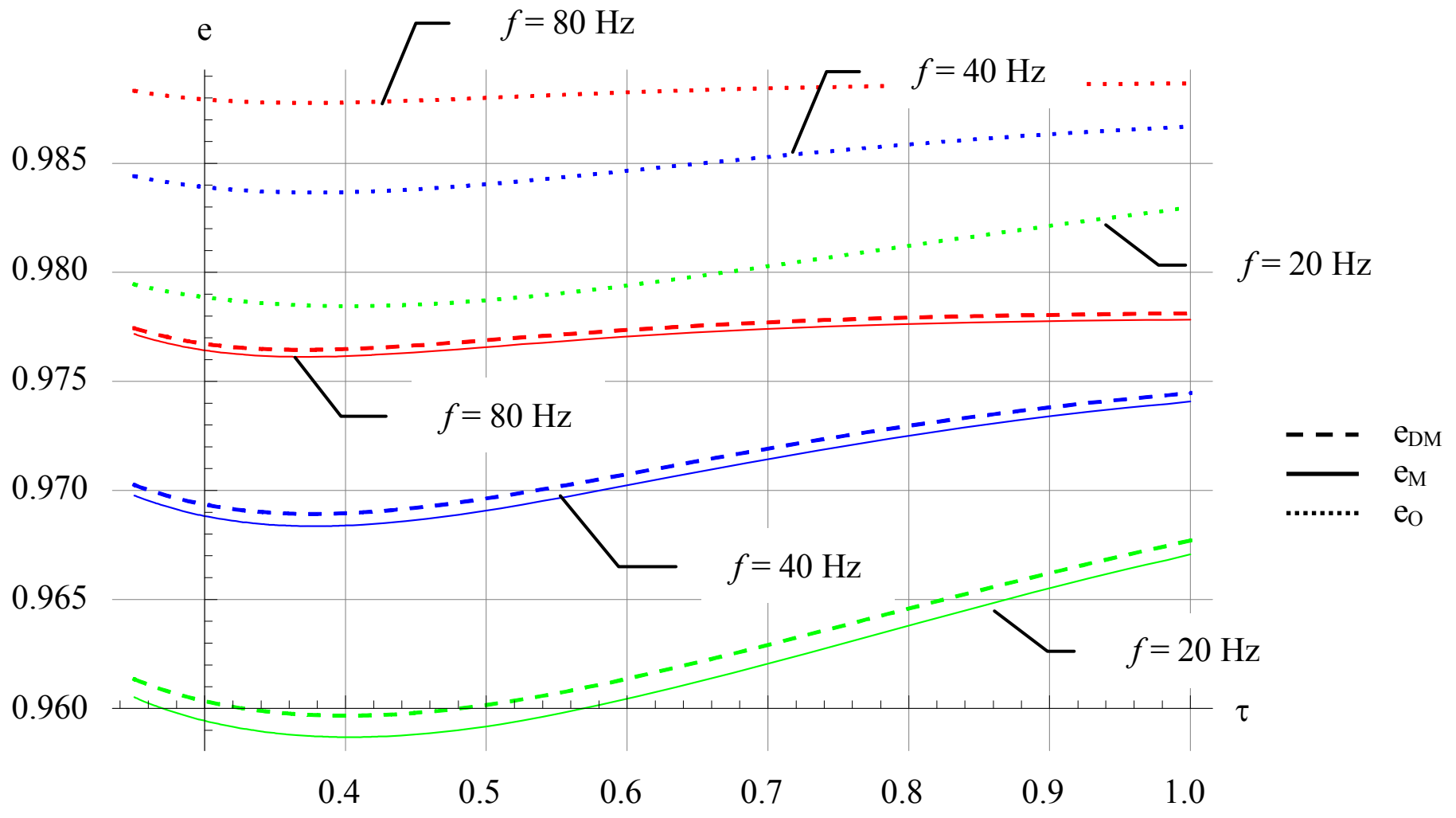

Fig. 5. Regenerator efficiency $e$ with respect to temperature ratio $\tau$ for three operating frequencies.

The Organ definition gives more important values than the two others. The flush ratio has little influence on the regenerator efficiency. As a consequence, the simple Martini expression will be used in the following.

- Conduction losses

Most of the thermal conduction appears through the regenerator which separates the hot and cold areas. We choose an expression of the heat flux related to the conduction losses based again on the Newton's law and a constant coefficient $c_{cond}$ defined as:

$$c_{cond} = k_w \, N_{conn} \frac{A_{ffreg}}{L_{reg}} \frac{1 - \P_v}{\P_v} \tag{25}$$

In which $N_{conn}$ is the apparent connectivity related to the thermal connexion in the mesh grid regenerator [19] and $k_w$ the thermal conduction of the regenerator grid material.

- Mechanical effectiveness

Save for thermal losses, mechanical losses must also be taken into account to evaluate the brake output power and global efficiency.

On the one hand friction occurs within kinematic linkages joints for the piston and displacer as well as for the output shaft, on the other hand moving parts seals can lead to great friction forces. Both count in a single mechanical effectiveness coefficient called $\eta_{mec}$. Moreover we assume a fixed value of $\eta_{mec}$ whatever the operating conditions.

Another source of mechanical dissipation lies in pressure drops associated to the fluid flow trough exchangers and regenerator denoted $\Delta p_{exch}$ and $\Delta p_R$ respectively. Pressure drops can be calculated using the friction factor coefficient related to the Reynolds number.

Therefore, the total mechanical effectiveness is:

$$\eta_{diss} = \eta_{mec} + \frac{\Delta p_{exch}}{p} + \frac{\Delta p_R}{p} \tag{26}$$

Senft has demonstrated the drastic effect of the buffer pressure on the mechanical effectiveness [22]. He established the central theorem as will be recalled hereafter.

The additional mechanical features are to be considered as shown on the schematic representation of Fig. 6. $\dot{W}_{out}$ is the brake output power and arrows stand for the power transfers between parts.

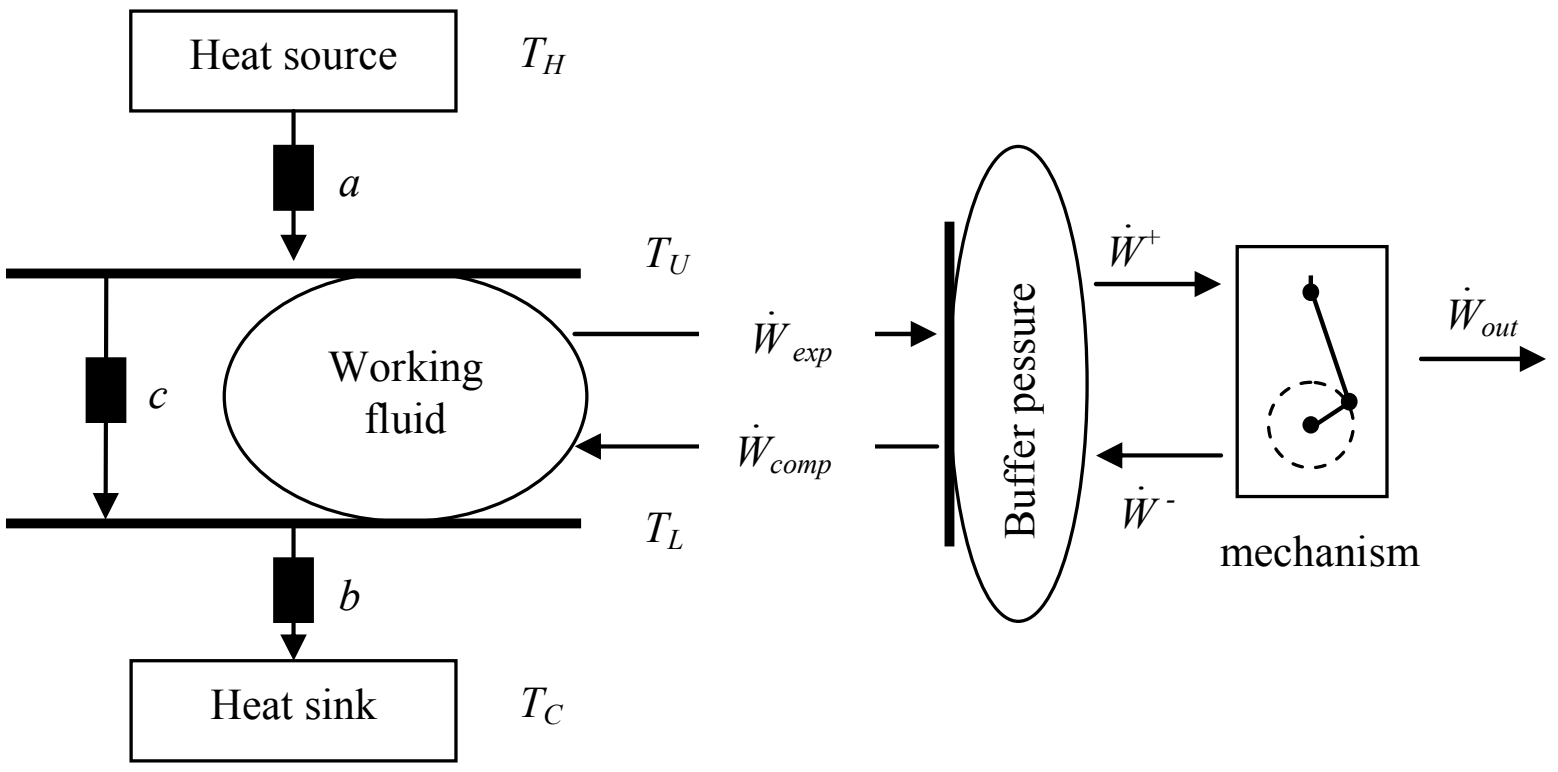

Fig. 6. General Stirling machine scheme with thermal interfaces buffer pressure and mechanical parts.

One can finally give the expression of the indicated power:

$$P_i = \dot{W}_i = \dot{W}_{exp} - \dot{W}_{comp} \tag{27}$$

And the output power:

$$P_{out} = \dot{W}_{out} = \eta_{diss} (\dot{W}^+ - \dot{W}^-) \tag{28}$$

The global final cyclic mechanical effectiveness $\eta_{MEC}$ is the ratio of the output power $P_{out}$ to the indicated power $P_i$.

The central theorem established by Senft is recalled hereafter:

*The cyclic mechanical effectiveness of any engine with volume compression ratio* $r = \frac{Vmax}{Vmin}$, *temperature ratio* $\tau = \frac{T_L}{T_U}$ *and mechanical effectiveness* $\varepsilon < E$, *cannot exceed* $\eta_{MEC}$:

$$\eta_{MEC}(E, \tau, r) = E - \frac{1 - E^2}{E} S(\tau, r)$$

$$S(\tau, r) = 0 \text{ if } \tau r \leq 1 \tag{29}$$

$$S(\tau, r) = \frac{\tau \log(\tau) - (1+\tau)(\log(1+\tau) - \log(1+r) - \log(r))}{(1-\tau)\log(r)} \text{ if } \tau r > 1$$

*This upper bound is the mechanical efficiency of an ideal Stirling engine buffered at its optimum constant pressure and having a mechanism of constant effectiveness E.*

Substituting $\eta_{diss}$ for $E$ in the above theorem, the cyclic mechanical efficiency $\eta_{MEC}$ can be added to the previous model. Fig. 7 plots the mechanical efficiency with respect to $\tau$. For the high temperature ratio of considered here, the effect of the buffer space pressure can be neglected. If low temperature differences are considered which is the case solar application the mechanical efficiency curve shows a sharp slope downward.

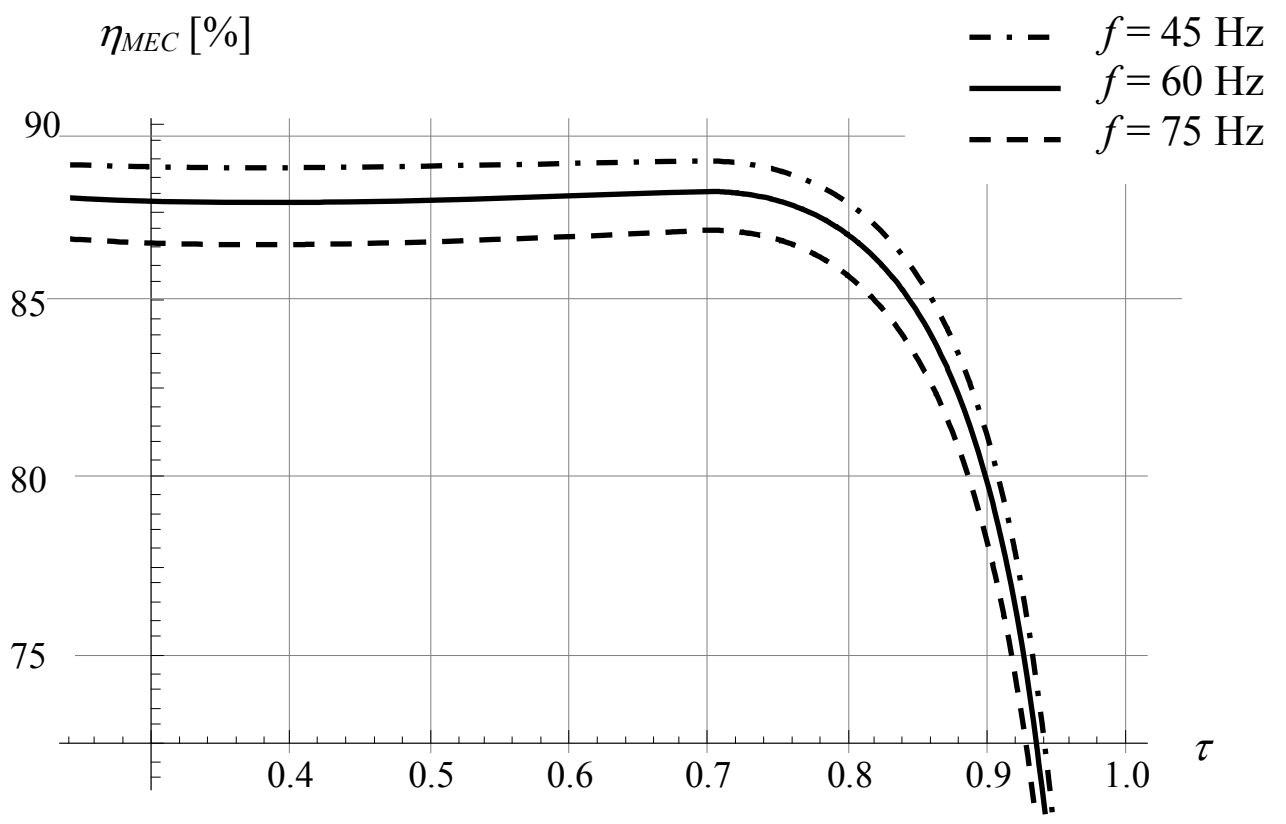

Fig. 7. Mechanical efficiency taking into account the buffer pressure effect.

### 4. Results and discussions

The operating point can be obtained from the equality of indicated and thermal power given in equations (11) and (12) respectively. Fig. 8 plots parametric curves of non-dimensional powers $P_i{*} = \dfrac{P_i}{f\,p_{mean}\,V_d}$ and $P_{th}^{*} = \dfrac{P_{th}}{f\,p_{mean}\,V_d}$ for various operating frequencies. The curves of the two powers with respect to temperature ratio $\tau$ intersect at one point.

Because of the enhanced heat exchange effectiveness, $P_{th}^{*}$ increases as $f$ increases and the operating point moves to lower values of $\tau$. However, if $f$ is greater than 90 Hz, the thermal transfer properties reach a limit resulting in an almost constant apex for $P_{th}^{*}$. The conclusions are quite different from the constant heat transfers case studied by Senft, in which the temperature ratio $\tau$ increases with $f$ in order to balance heat transfer governed by the Newton’s law. Therefore, the temperature differences between external and internal temperatures always increase.

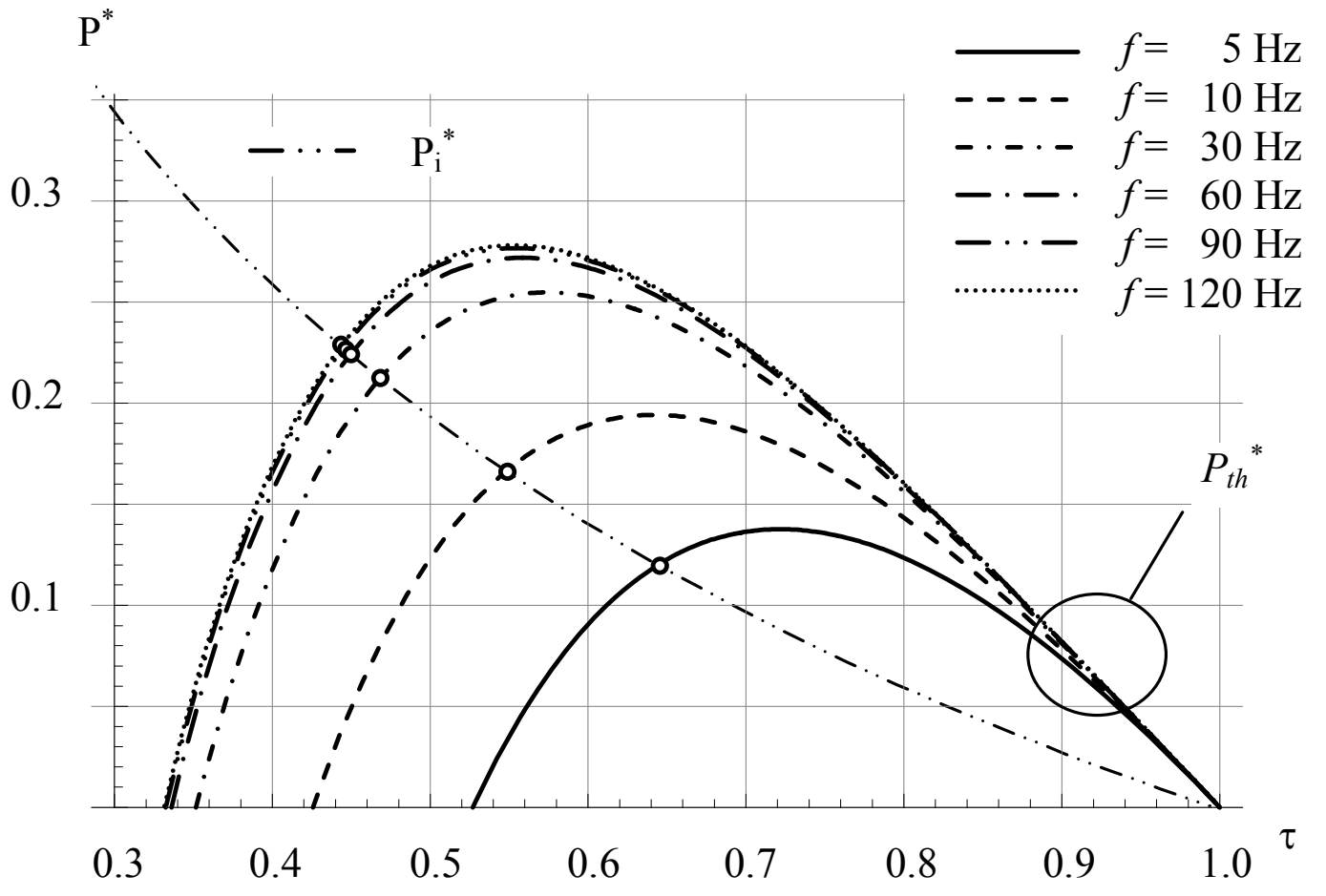

Fig. 8. Evolution of the operating point for various frequencies.

### 4.1. Model validation

The model results are compared with those obtained by Urieli and Berchowitz [9] and Timoumi [14] for the same conditions of the GPU-3 (*i.e.* working fluid Helium, mean pressure $p_{mean}$ = 4.13 MPa, and operating frequency $f$ = 41.72 Hz). The results presented in Table1 shows a good agreement with adiabatic models which take into account the pressure drop, the regenerator efficiency and conduction losses.

| Type of model | Heat input | Indicated power output | | Thermal efficiency |
|---|---|---|---|---|
| | J/cycle | W | J/cycle | % |
| Adiabatic model | 327 | 8286.7 | 198.62 | 62.06 |
| Urieli and Berchowitz [9] adiabatic model | - | 8300 | - | 62.5 |
| Timoumi dynamic model without losses [14] | 314 | 7109.3 | 170.4 | 54.96 |
| Urieli and Berchowitz quasi-steady flow | - | 7400 | - | 53.1 |
| Timoumi dynamic model with loss dissipation (M1) | 291 | 6372.4 | 152.47 | 53.3 |
| Urieli and Berchowitz quasi-steady flow (pressure drop included) | - | 6700 | - | 52.5 |
| Timoumi dynamic model M1 and internal conduction loss (M2) | 294 | 6355.2 | 152.32 | 52.64 |
| **Developped analytical isothermal model** | **258** | **6087** | **146** | **52.9** |
| Timoumi (M2) and external conduction loss | 314 | 6061 | 145.27 | 46.94 |
| Timoumi dynamic best model | 262 | 4273 | 99.5 | 38.49 |
| Experiment | - | 3958 | - | 35 |

Table 1. Results from different models.

As expected from the isothermal assumption, the predicted efficiency is 13% higher and the power and the work by cycle are very close when compared to the adiabatic model without shuttle and gas spring hysteresis losses.

Most of the studies deal with a single operating point for the validation stage. Because the model is dedicated at design and optimization, it is compared hereafter to GPU-3 experimental values for a wide range of parameters especially for the pressure and frequency.

The geometrical parameters as well as operating data for General Motor GPU-3 are summarized in Table 2 below.

| | | | | | |
|---|---|---|---|---|---|
| Heat source $T_H$ | [°C] | 900 | Cold source $T_C$ | [°C] | 15 |
| Phase angle $\alpha$ | [deg] | 120 | *swept volume ratio K* | [ - ] | 1.01 |
| Swept volume $V_d$ | [cm$^3$] | 120.88 | Working fluid | | Hydrogen |
| Mean pressure $p_{mean}$ | [MPa] | $0.69 < p < 6.89$ | mechanical effectiveness $\eta_{mec}$ | [%] | 90 |
| Operating frequency $f$ | [Hz] | $20 < f < 50$ | | | |
| Wire mesh regenerator | | | regenerator length | [mm] | 22.6 |
| wire diameter $d_w$ | [µm] | 40 | porosity of the regenerator matrix ¶$_v$ | | 0.759 |
| regenerator hydraulic radius $r_{hR}$ | [mm] | 0.03 | Material conductivity | [Wm$^{-1}$K$^{-1}$] | 16.6 |

Table 2. Parameter values of the GPU-3 Stirling engine.

To allow easier comparison, the performances of the model are expressed in original units from the GPU-3 experimental test results [15]. The Fig. 9 shows the variation of the torque, brake output power and efficiency with respect to operating frequency for various mean pressures. The dotted lines are from the experimental results whereas the straight lines stand for the model results.

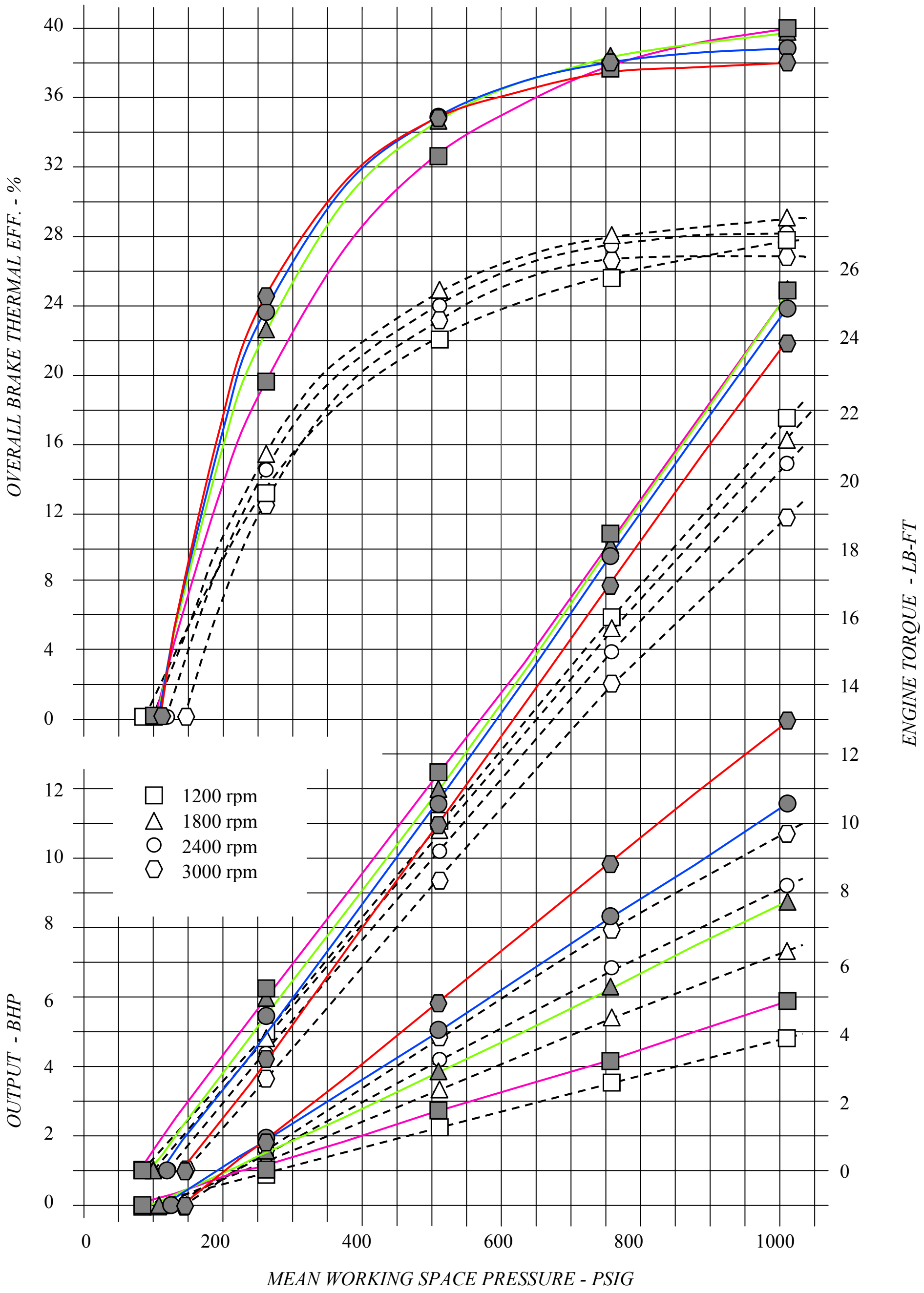

Fig. 9 Brake output power and efficiency curves with respect to operating frequency and pressure.

The model shows good agreement with the whole data. The analytical model qualitatively represents the trends of the curves. Due to the isothermal assumption, the efficiency from the model is above the experimental one as expected from this type of models [15, 23]. The efficiency results show a constant discrepancy of about 46% with the experimental ones.

The experimental results show a static torque which prevents the engine from running for the lowest value of the mean pressure. This effect is not taken into account in the model. Consequently, the torque results are forced to zero by subtracting the value obtained for the lowest mean pressure. By doing this for the torque and the power, a constant difference of about 25 % for each frequency is obtained. It is worthy of note that the reduction of the torques values by increasing the operating frequency is well reproduced by the model.

## 4.2. Optimization

### 4.2.1. Effect of the fluid mass

The total mass of the gas is related to the mean pressure by the equation (6). The proportional relation between the power and the mean pressure can be seen in Fig. 10 a.

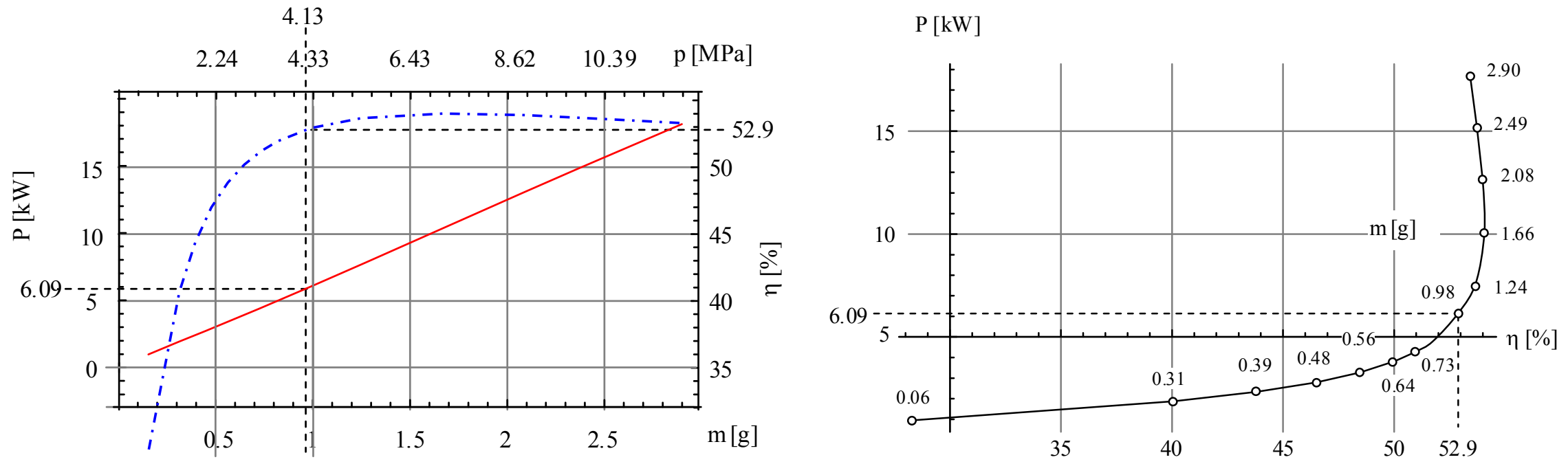

(a) (b)

Fig. 10. Influence of the fluid mass brake output power and efficiency.

This result is close to the results obtained for the GPU-3 and the effect of the fluid mass on the efficiency is different for the work of Timoumi [14]. An optimal fluid mass of $m$ = 1.6 grams (i.e. a mean pressure $p_{mean}$ = 6.89 MPa) for which a maximal value of 54% efficiency can be inferred (see Fig. 12 b) which is 2% higher than for the prototype reference value shown in dashed line in Fig. 10. In the same time, the power is raised from 6.09 kW to 10 kW.

#### 4.2.2. Combined effect of fluid mass and frequency

For a given mass of gas, the operating frequency can be optimized to improve the performances. Fig. 13 shows the evolution of the power-efficiency curves for various pressures. The GPU-3 reference operating value of 41.72 Hz appears to be close to the optimal value whatever the pressure.

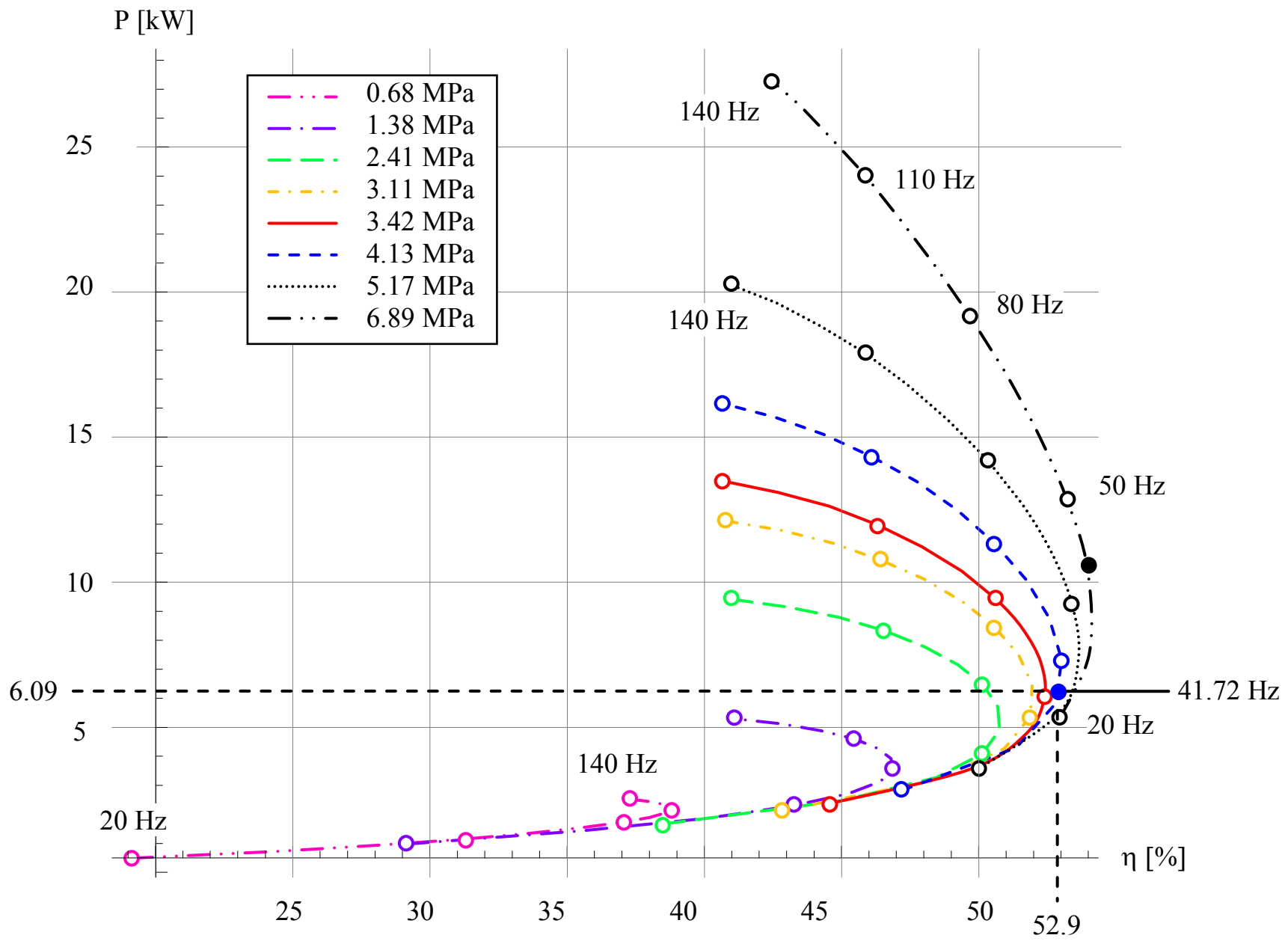

Fig. 11. Influence of the mean pressure on the brake output power – efficiency curve.

#### 4.2.3. Effect of the regenerator length

By varying the regenerator length, the performances are modified accordingly. The additional dead volume as well as an increase of the pressure loss make the power reaches a maximal value. However, for small regenerator length *i.e.* below 15 mm, the power as well as efficiency quickly decreases, as shown in Fig. 12 a.

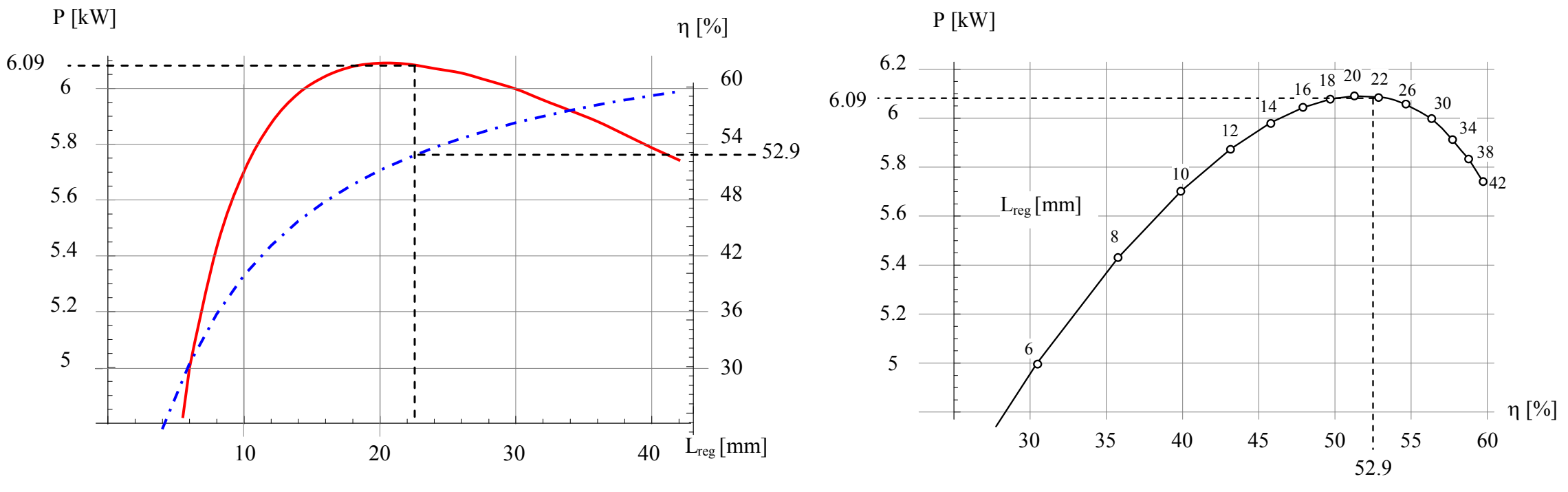

(a) (b)

Fig. 12. Influence of the regenerator length on the brake output power and efficiency.

The power-efficiency curve shown in Fig. 12 b underlines the choice of the optimal power optimization chosen for the GPU-3 operating parameters.

### 4.2.4. Regenerator thermal conductivity

The conductivity of the material constituting the regenerator matrix has strong effects. Fig. 13 shows that with an increase of the matrix regenerator thermal conductivity leads to a reduction of the performance due to an increase of internal conduction losses in the regenerator. For a value close to the steel thermal conductivity ($k_w \approx 40$ $Wm^{-1}K^{-1}$), the optimal efficiency is about 19 % at a higher frequency than for the stainless steel used for the GPU-3. For a low conductivity material such as the Inconel 625® ($k_w \approx 9.8$ $Wm^{-1}K^{-1}$), the efficiency can be increased to 57%.

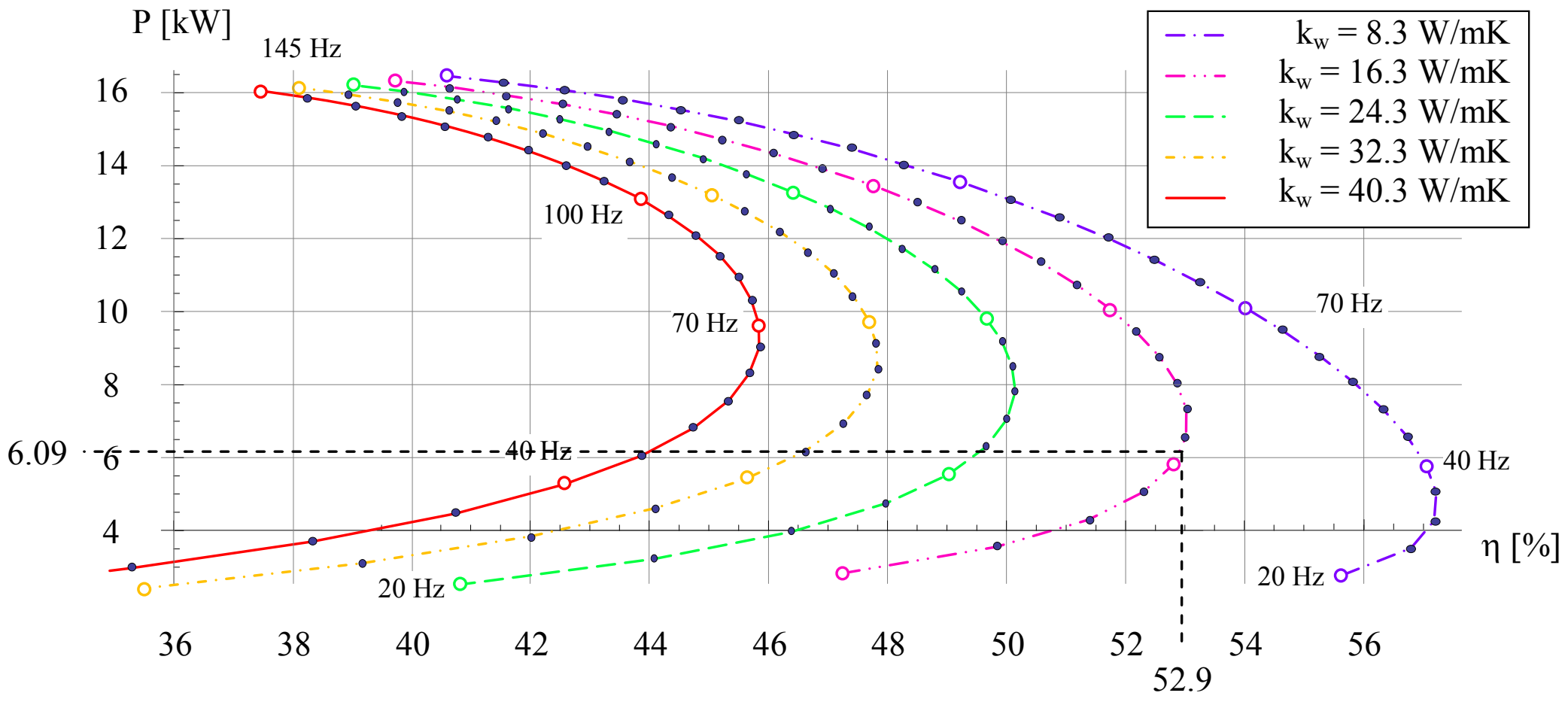

Fig. 13. Influence of the conduction loss on the brake output power – efficiency curve.

#### 4.2.5. Effect of the cooler efficiency

Although water cooling is used for the GPU-3 engine, a Stirling engine can use free or forced air flow for the cooler. Therefore for technological design, one can be interested in cold heat transfer coefficient appraisal.

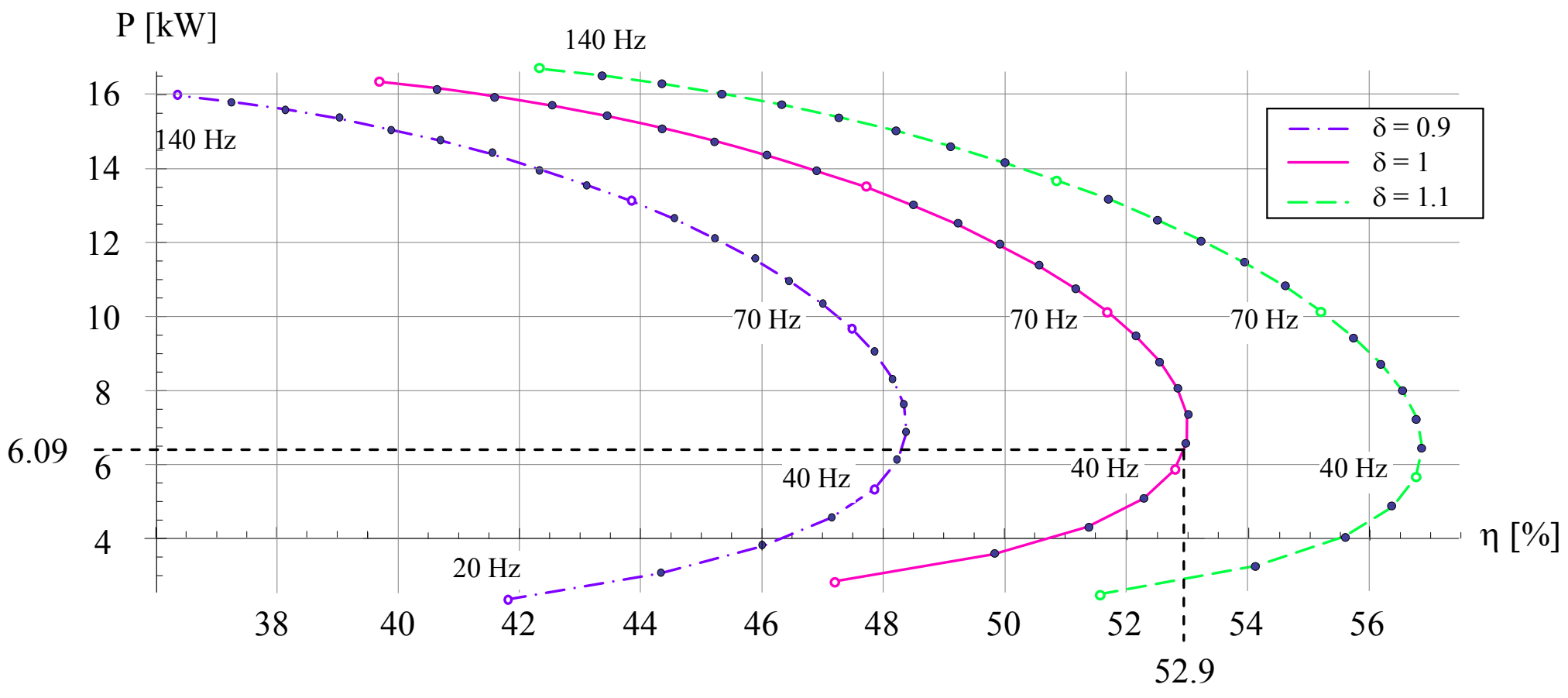

Fig. 14. Influence of the cold heat exchanger performance on the brake output power – efficiency curve.

An increase of 10% of δ leads to a sensitive variation of both the efficiency for a given frequency. Conversely, a decrease of 10% induces the same absolute difference on the efficiency. Again, the major role of heat transfers in Stirling engine is outlined here.

## *5. Conclusion*

As a first step for design and optimization of FPSE, an analytical model has been elaborated. To deal with the discrepancy between the high theoretical efficiency of Stirling and a constructed prototypes the major losses are be taken into account. Moreover, they are related to the geometrical and physical parameters of the prototype design.

In order to obtain a simplified analytical model, an isothermal assumption is adopted. The model integrates the regenerator efficiency and conduction losses, the pressure drops and the additional heat exchangers effectiveness. Thus, the influence of the operating parameters on the global engine performance can be assessed. As a validation stage, we applied the parameters of the GPU-3 engine data on the developed model. Despite using the isothermal assumption and simplified heat transfer model, the results are close to the experimental data and are in agreement with classical models.

An optimization of these parameters has been carried out using the GPU-3 engine data. It is shown that a reduction of losses and a notable improvement in the engine performance can be reached. In addition to the well known influence of the regenerator, the drastic effect of the cooler effectiveness is underlined.

As a conclusion, a handy model is developed and can be used in a first design procedure and optimization with respect to a specific application.